\documentclass[letterpaper, 10 pt, conference]{ieeeconf}
\IEEEoverridecommandlockouts
\usepackage{mathptmx} 
\usepackage{times} 
\usepackage{amsmath} 
\usepackage{amssymb}  
\usepackage{amsfonts}
\usepackage{xspace}
 \usepackage{epsfig}
\usepackage{graphicx}
\usepackage{color}
\usepackage{soul}
\usepackage{subfigure}

\newcommand{\gamovertwo} {{\frac{\gamma^2}{2}}}
\newcommand{\gammovertwo} \gamovertwo
\newcommand{\gammuovertwo} \gamovertwo
\newcommand{\gammuepstovertwo} \gamovertwo
\newcommand{\gammubarepstovertwo} \gamovertwo

\newcommand{\noncr} {\nonumber\\}

\newcommand{\beasnum}{\begin{eqnarray}}
\newcommand{\eeasnum}{\end{eqnarray}}
\newcommand{\beas}{\begin{eqnarray*}}
\newcommand{\eeas}{\end{eqnarray*}}

\newcommand{\be}{\begin{equation}}
\newcommand{\ee}{\end{equation}}
\newcommand{\ba}{\begin{array}}
\newcommand{\ea}{\end{array}}

\newcommand{\ket}[1]{|{#1}\rangle}

\newcommand {\dpd}[2] {\frac{{\partial}^{2} {#1}}{\partial {#2}^{2}}}
\newcommand {\spd}[2]{\frac{\partial {#1}}{\partial {#2}}}

\def \secname {sec:optqstatecdc:2011}
\def \eqnname {eqn:optqstatecdc:2011}

\def  \figname {fig:optqstatecdc:2011}

\def \solnot {\theta}
\def \aeq {&=}
\def \adefeq {&:=}

\def \notncsigval {v}

\newcounter{AssumptionCounter}
\stepcounter{AssumptionCounter}

\newcommand {\controlSet} {\mathcal{V}}
\newcommand {\controlValueSet} {\mathbf{V}}

\def \controlBound {\Omega}

\def \targetSet {\mathcal{T}}

\def \thetaset {G}
\newcommand{\thn}[1] {\theta_{{#1}}}
\def \eop {\mathrm{E}}
\newcommand {\Eop}[1]{\mathrm{E}\big[{#1}\big]}

\def \ctrlboundsimval {5}
\def \betasimval {0.1}
\def \gammasimLim {1}
\def \rotationsigSet {\controlSet}
\def \meassigSet {\Im}
\def \cpair {\eta}
\def \cpairsigSet {\Xi}

\begin{document}
\title{Optimal rotation of a qubit under  dynamic measurement and velocity control}
\author{Srinivas Sridharan \thanks{}}
\author{
\authorblockN{Srinivas Sridharan}
\authorblockA{Dept. of Mechanical and Aerospace Engineering \\
University of California San Diego \\
Email: {srsridharan@ucsd.edu}}
\thanks{Research supported by  AFOSR grant FA9550-10-1-0233.}
}

\maketitle
\begin{abstract}
In this article we explore a modification in the  problem of controlling the rotation of a two level quantum system from an initial state to a final state in minimum time. Specifically we consider the case where the qubit is being weakly monitored -- albeit with an assumption that both the measurement strength as well as the angular velocity are assumed to be control signals. This modification  alters the dynamics significantly and enables the exploitation of the measurement backaction to assist in achieving the control objective.  The proposed method yields a significant speedup in achieving the desired state transfer compared to previous approaches. These results are demonstrated via numerical solutions for  an example problem on a single qubit.

\end{abstract}
\section{Introduction}\label{\secname intro}

One of the requirements common to many applications of quantum engineering \cite{nielsen2000qca,higgins2007entanglement,WisMil10}, is that of being able to create a quantum system in a particular pure quantum state. Moreover, it is desirable to achieve such a state value in as short a time as possible. This has motivated 
 research into this domain of time optimal control of quantum systems; 
Although has been a significant amount of research on optimal control for closed quantum systems \cite{dalessandro2002ufg,schirmer2001qcu,ramakrishna2000qcd, nielsen2006qcg,N.2001}, the investigation into  the optimal control of monitored open quantum systems  still remains to be developed fully \cite{belavkin1999measurement, wiseman2008optimality,shabani2008locally,belavkin2009dynamical,gough2005hamilton}.

 In this work we address the problem of transferring the state of a two level quantum system from an 
 initial pure state  to a final pure state. In particular develop the idea proposed in \cite{sjm2011optimalqbitcontrol} on the possible use of  measurement backaction for speeding up the control of a qubit to a greater extent that that achievable by fixed measurement alone. The framework for this 
 problem is unique due to the fact that both the angular rotation as well as the 
 measurement strength for the continuous measurement  being performed are control signals that can be varied to achieve the desired objective. This leads to a significant speedup in the time required to reorient the system between two pure states. Intuitively we use the fundamental property of measurement backaction, unique to quantum systems, to help achieve the state transition more rapidly than that possible  using either: static measurement and feedback  or that with no-measurement and a control driven by a constant rotation (termed the Hamiltonian evolution).

The structure of this article is as follows. 
In section~\ref{\secname problem}  we describe the system model and frame the optimal control problem of interest. 
 To solve the problem described above, we utilize the dynamic programming approach from optimal control theory in section~\ref{\secname ochittingtime}.  It turns out however, that the Hamilton-Jacobi-Bellman  equations associated with this control problem may not have classical solutions. We discuss this issue and indicate a numerical method to obtain the solution in section~\ref{\secname ochittingtime}. 
 Two cases of interest in the optimal control of the state of a quantum system are transition between states which are:  (a)   parallel or (b) orthogonal to the axis of measurement.
The solution to the control problem for these different cases are described in section~\ref{\secname examples}.  We then highlight the performance of the dynamic measurement and control strategy introduced herein with respect to that achieved using either only static measurement strength or with pure rotation alone (with no measurement).
We conclude with  a discussion of further interesting questions that arise from these results, in section \ref{\secname conc}.

\section{System  Description and further background}\label{\secname problem}
\subsection{The model}
We now explain the model for the system of interest - the continuous measurement and feedback control of a single qubit.
{A more comprehensive introduction to continuous quantum measurements can be found in \cite{brun2002,steck2006}.} 
The state of such a system may be represented  by a vector in 
a $3$~dimensional real unit sphere (termed the \emph{Bloch sphere}). 

Consider a quantum spin system subject to measurement along the $z$ axis\footnote{i.e., this measurement is of the observable $\sigma_z$.}, with  control signals that induce a rotation around the three axis. Let $\gamma$ denote the measurement strength -- a parameter that determines the rate 
at which the information is extracted from a system. A larger value of $\gamma$ leads to a greater rate 
of information extraction  and therefore a higher rate at which the system is projected on to one of the eigenstates of the observable used [in this case it is the Pauli matrix $\sigma_z = \mathrm{diag}(1,-1)$]. For the particular choice of measurement considered herein, 
these  eigenstates correspond to the up ($+z$) direction and the down ($-z$) direction on the Bloch sphere.
The goal of the feedback is to take the initial state (say $\ket{0}$ i.e., the $+z$ direction) and control it to the orthogonal state (say $\ket{1}$ i.e., the $-z$ direction) in a time optimal manner. 
We assume that: 
\begin{enumerate}
\item  the initial state is  pure and that measurements are efficient. This implies that the 
evolution of the state vector is confined to the surface of the Bloch sphere. 
\item the available control is equal in strength (isotropic) about all 3 axes. 
\end{enumerate}
Under the above assumptions, by using the symmetry of the problem the Bloch representation  reduces (after a change of variables) to  a lower dimensional model;  Herein the state is constrained to lie on the unit circle i.e., the $x$-$z$ plane. The control signal causes the state to rotate around the $y$ axis. The control problem for this system involves moving the state between any two states on the unit circle. 

The system described above is  modeled using a stochastic differential equation (SDE) of the form 
\begin{align}
d\theta(t) \aeq \alpha(t)\, dt - 2 \gamma(t) \sin(2 \theta(t)) dt - 2 \sqrt{2 \gamma(t)} \sin(\theta(t)) dW. \label{\eqnname syseqn}
\end{align}
where $\theta \in[-\pi,  \pi]$.
The term  $\alpha$ denotes the control signal applied. To
 ensure that the control problem is well posed we apply a bounded strength control, i.e., the 
controls are constrained  to a  closed compact set $\controlValueSet:=[-\controlBound, \controlBound]$. Here the maximum and minimum values that the control signal can take up at any time instant are symmetric and have a magnitude $\controlBound$.
We  denote the set of piecewise continuous angular velocity signals by the term $\rotationsigSet$.

 The second term  in the equation above is the quantum measurement backaction. This can be intuitively understood by setting the angle to $0$ or $\pm \pi$, so the state vector and the measurement axis commute and the backaction goes to zero. Similarly at the point  $\theta = \pm \pi/2$ where the state and measurement axes are maximally non-commuting, the measurement back-action is largest.
The term $\gamma (t)$ denotes the measurement strength which is also a control term in this problem. The values allowed for the measurement strength belong to the 
range $[0, \Gamma]$. The set of measurement strength signals is denoted by $\meassigSet$.  The final term in Eq.~\eqref{\eqnname syseqn} indicates the innovation term arising from measurements. 
 
We denote the control signal pair $(\alpha, \gamma)$ as $\cpair$ and the set from which this signal is drawn as $\cpairsigSet := \rotationsigSet \times \meassigSet$. 
The solution to the SDE \eqref{\eqnname syseqn} for a trajectory, starting at a point $\thn{0}$ (at a time $t_{0}$) and using a control {strategy} $\cpair \in \cpairsigSet$, at a time  instant $t\in [t_{0},\infty)$
is denoted by $\solnot(t;\cpair,t_{0},\thn{0})$. Note that the this trajectory is a particular sample path of a stochastic process. In cases where the  arguments used in this notation are clear from the context we represent the solution at time $t$ by  $\thn{t}$. 

\begin{figure}
\begin{center}
\leavevmode \includegraphics[width=0.5\hsize]{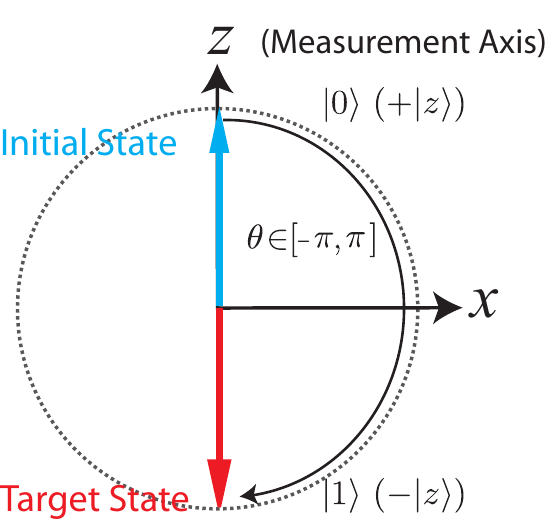}
\caption{The Bloch sphere with a graphical depiction of our control problem. We start in the plus eigenstate of the observable $\sigma_{z}$ and rotate to the orthogonal state $-\ket{z}$. \label{\figname Fig1} }
\end{center}
\end{figure}

\subsection{Performance measure: the expected (discounted) hitting time}\label{\secname costfnz}
In order to capture the time optimality requirement of the problem, the most intuitive approach is to model  the problem as an optimal control problem with a cost function that measures the expected hitting time to the target set (denoted by $\targetSet$).  Define the hitting time as
\begin{align}
\tau^{\cpair}_{\targetSet}(\thn{0}) \aeq \inf\{t | \solnot(t; \cpair,t_{0},\thn{0}) \in \targetSet\}\label{\eqnname hittingtime},
\end{align}
i.e., for each sample path it is the first time at which the trajectory reaches the target set (and is a random variable).  
The cost function based on the expected hitting time  will take the form
\begin{align}
{C}(\thn{0}) \adefeq \inf_{\cpair\in \cpairsigSet}\Eop{\tau^{\cpair}_{\targetSet}(\thn{0})},\\
\aeq \inf_{\cpair\in \cpairsigSet}\eop\Big\{\int_{0}^{\tau^{\cpair}_{\targetSet}(\thn{0})}{ds}\Big\}.
\end{align}

However, as described in the next section, application of the dynamic programming principle to solve this optimal control problem  leads to an associated  Hamilton-Jacobi-Bellman (HJB) equation whose classical solution may not exist. 
Thus we use a revised discounted  cost function  which ensures the uniqueness a weak solution in similar classes of optimal control problems (c.f. \cite{fleming2006cmp,sridharan2010numerical}) \footnote{We defer the discussion of the weak solutions and a rigorous justification of the discounted cost in this problem to the future.}
\begin{align}
C(\thn{0}) \adefeq \inf_{\cpair\in \cpairsigSet}\eop\Big\{\int_{0}^{\tau^{\cpair}_{\targetSet}(\thn{0})}{\exp\{-\beta s\}ds}\Big\}\label{\eqnname discountedcostfn}.
\end{align}
The parameter $\beta > 0$ is called the discount factor. One  interesting aspect  of this  discounting is that  the cost function remains bounded for any choice of the control signal.
The motivation for and advantages of this discounting will be discussed in detail below.

\section{Optimal control for the hitting time problem}\label{\secname ochittingtime}
In order to obtain the optimal cost function (Eq.~\eqref{\eqnname discountedcostfn}) and the corresponding control strategy for the system of interest, we apply the dynamic programming \cite{bellman2003dp,bertsekas1995dpa} approach from optimal control theory. 

Note that the system dynamics in Eq.~\eqref{\eqnname syseqn} is an SDE  of the form
\begin{align}
dx = {\bf b}(x,\alpha,\gamma)dt + {\boldsymbol \sigma}(x,\gamma) dW.
\end{align}
By comparison, the  coefficients ${\bf b}, {\boldsymbol \sigma}$  can be seen to be 
\begin{align}
{\bf b}(x, \alpha,\gamma) \adefeq  \alpha - 2 \gamma \sin(2 \theta),\\
{\boldsymbol\sigma}(x, \gamma) \adefeq 2 \sqrt{2 \gamma} \sin(\theta).
\end{align}

We introduce a differential operator $L^{\notncsigval}[\phi](y)$  given by the expression
 \begin{align}
 L^{\cpair}[\phi ] (y) \adefeq  \mathbf{b}(y, \cpair) \spd{\phi}{\theta} \Bigg |_{\theta = y}+ \frac{1}{2} {\boldsymbol\sigma}^{2}(y,\gamma) \dpd{\phi}{\theta}\Bigg |_{\theta = y} \label{\eqnname loperatordefn},
\end{align}
which is  the generator of the It\={o} diffusion process Eq.\eqref{\eqnname syseqn}. 
The application of dynamic programming  to this optimal control problem yields  the following Hamilton-Jacobi-Bellman equation over the set $G:= (-\pi, \pi)$:
\begin{align}
\sup_{\cpair\in \cpairsigSet}  \left \{-1+ \beta \phi - L^{\cpair}[\phi ] (y) \right\} =0, \quad \forall y \in \thetaset \label{\eqnname hjbdef1}
\end{align}
with boundary conditions
\begin{align}
\phi(\targetSet_{e}) \aeq 0.
\end{align}
The classical solution to this partial differential equation (PDE) yields the  discounted cost function in Eq.~\eqref{\eqnname discountedcostfn}. 

 Note that the HJB equation \eqref{\eqnname hjbdef1} is an elliptic PDE \cite{gilbarg2001elliptic} with a coefficient for the second order derivative that can become zero at any point in the domain $G$ -- therefore it is called a degenerate elliptic PDE. 
%
The positivity (non-degeneracy) of this second order term is a sufficient condition for the existence of a classical solution to this PDE \cite{wong1985stochastic,fleming2006cmp}. Hence, due to the nature 
of the $\sigma(\cdot,\cdot)$ term in the system dynamics being able to take up a value of zero, there arises a degeneracy owing to which the HJB equation is not guaranteed to have a sufficiently smooth solution. Therefore a rigorous study of the
solution to the optimal control problem necessitates an analysis of the solution to this equation in a weak sense.

It is interesting to note that an alternate approach  used in the literature to determine the hitting time involves  solving a PDE termed the Fokker-Planck equation \cite{gardiner1985handbook, jacobs2010stochastic} . The solution to this equation is the probability density of the distribution of the hitting time (from which we can evaluate the expectation of the hitting time). The degeneracy indicated above also arises naturally in the Fokker-Planck equation, thereby giving rise to the same issues  of non-existence of classical solutions.

In this article our focus is on obtaining and analyzing the optimal control strategy and the improvement obtained in the time optimality in state transfer compared to that achieved by other strategies.
Hence we defer the analysis of these questions of existence and uniqueness of the generalized solutions for this problem to a future publication. 

\subsection{Numerical solution}
One widely applicable method for computing the solution to optimal control problems is the Markov chain approximation method \cite{kushner1992nms,fleming2006cmp}. In this approach the system dynamics are approximated by a controlled Markov chain on a finite state space. The cost function is then approximated by a discretization suited to this chain. Thus an iteration is constructed which converges to the desired cost function under the limit that the discretization converges towards the original formulation.  For a more detailed introduction to this approach and other applications to quantum control we refer the reader to \cite{kushner1992nms, smjpra2008,sridharan2010numerical} and the references therein. 
We now outline the iterative procedure to solve for the cost function.

Define 
\begin{align}
a^{+}\adefeq \max\{a, 0\}, \quad a^{-} := \max\{-a, 0\}.
\end{align}
We denote the spatial discretization interval by \lq$h$\rq. 
In addition, we use the expression 
\begin{align}
\exp\{- \beta \Delta t\} \approx \frac{1}{1 + \beta \Delta t},
\end{align}
to approximate the exponential weighting term. 
We generate a grid ${G}^{h}$ that approximates the set $G$ (for instance using a mesh with step-size $h$). 
The discretization for the HJB equation \eqref{\eqnname hjbdef1}  yields
\begin{align}
\phi^{h}(x) = \min_{\cpair\in \cpairsigSet} \Bigg\{ \Big[\sum_{y} p^{h}(x,y|\cpair) \phi^{h}(y) + \Delta t^{h}(x, \cpair)\Big] \times \noncr \frac{1}{1 + \beta \Delta t^{h}(x,\cpair)} \Bigg\}\label{\eqnname dischjb},  \quad x \in \tilde{G}
\end{align}
where the summation is over all points $y$ neighboring $x$.   The terms \lq $p$\rq\, in the equation   above are functions that are  given by \cite{kushner1992nms}. 
\begin{align}
p^{h}(x, x+h | \cpair) \adefeq \frac{\sigma^{2}(x,\gamma)/2 + h b^{+}(x,\alpha,\gamma)}{{\sigma^{*}}^{2}(x) + h B^{*}(x)},\\
p^{h}(x, x-h | \cpair) \adefeq \frac{\sigma^{2}(x,\gamma)/2 + h b^{-}(x,\alpha,\gamma)}{{\sigma^{*}}^{2}(x) + h B^{*}(x)},\\
p^{h}(x, x| \cpair) \adefeq \frac{[{\sigma^{*}}^{2}(x) - \sigma^{2}(x,\gamma)]+ h B^{*}(x) - h |b(x,\alpha,\gamma)|}{{\sigma^{*}}^{2}(x) + h B^{*}(x)},
\end{align} 
where
\begin{align}
B^{*}(x) \adefeq \Omega + 2 \Gamma |sin(2x)|, \quad \sigma^{*}(x) := 2 \sqrt{2 \Gamma} sin(\theta), \\
\Delta t^{h}(x, \cpair) \adefeq \frac{h^{2}}{{\sigma^{*}}^{2}(x)+ h B^{*}(x)}.
\end{align}

Denoting the RHS of Eq.~\ref{\eqnname dischjb} as an operator $\xi$ acting on the value function $\phi(\cdot)$ we obtain the iteration 
\begin{align}
\phi_{k+1}^{h}(x) = \xi(\phi_{k}^{h})(x), \quad x \in G^{h},
\end{align}
Under appropriate choices of parameters, this operator $\xi$ can be shown to be a contraction mapping, thereby yielding the necessary convergence. 
The results obtained by applying this iterative method  to the problem of interest will be described in the following section.

\begin{figure}[htp]
\begin{center}
\leavevmode \includegraphics[width=0.6\hsize]{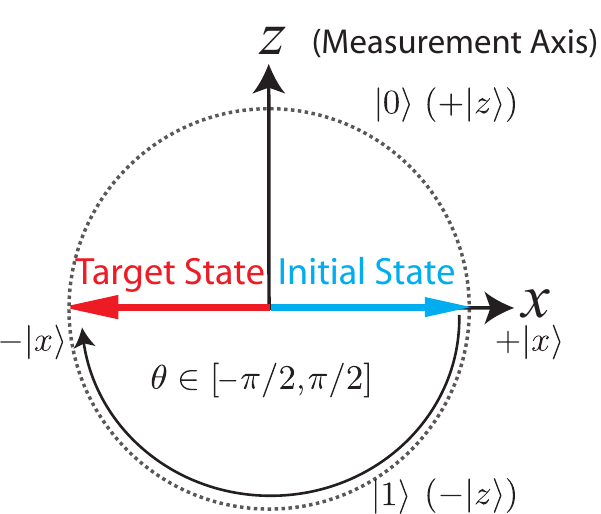}
\caption{The Bloch sphere with a graphical depiction of our control problem. We start in the plus eigenstate of the observable $\sigma_{x}$ and rotate to the orthogonal state $-\ket{x}$. \label{\figname Fignoneigtononeig} }
\end{center}
\end{figure}

 \section{Numerical Examples}  \label{\secname examples}
In this section we describe the solution to the optimal control problem for two cases of interest: 
\begin{enumerate}
\item  when the initial and final states for the control problem are
eigenstates of the observable $\sigma_z$ i.e., to move from $\ket{0}$ to $\ket{1}$ (ref. Fig.~\ref{\figname Fig1}). 
\item when the states are both maximally non-commuting with respect to the measurement (ref. Fig.~\ref{\figname Fignoneigtononeig}) with $\sigma_z$ i.e., 
the problem is to go from $+\ket{x}$ to $-\ket{x}$. 
\end{enumerate}
These two cases are of interest since they help  clarify whether  the control problem between two orthogonal states depends on the nature of the initial and terminal points. The optimal control 
problems are solved numerically via the value iteration approach described in the previous section. 

In order to implement this approach we include a stopping criteria for the value iteration algorithm, which in our approach  is obtained from a stopping test function -- the maximum absolute value of the change in the cost function over all grid points. Once this goes below a fixed threshold, we stop the value iteration. Note that this is possible due to the fact that the value iteration operator $\xi$ is a contraction mapping (if not, there  would be no reason for this stopping test function to remain below the threshold in subsequent operations).
\subsection{Optimal transition between eigenstates}
\begin{figure}
\begin{center}
\includegraphics[width=0.9\hsize]{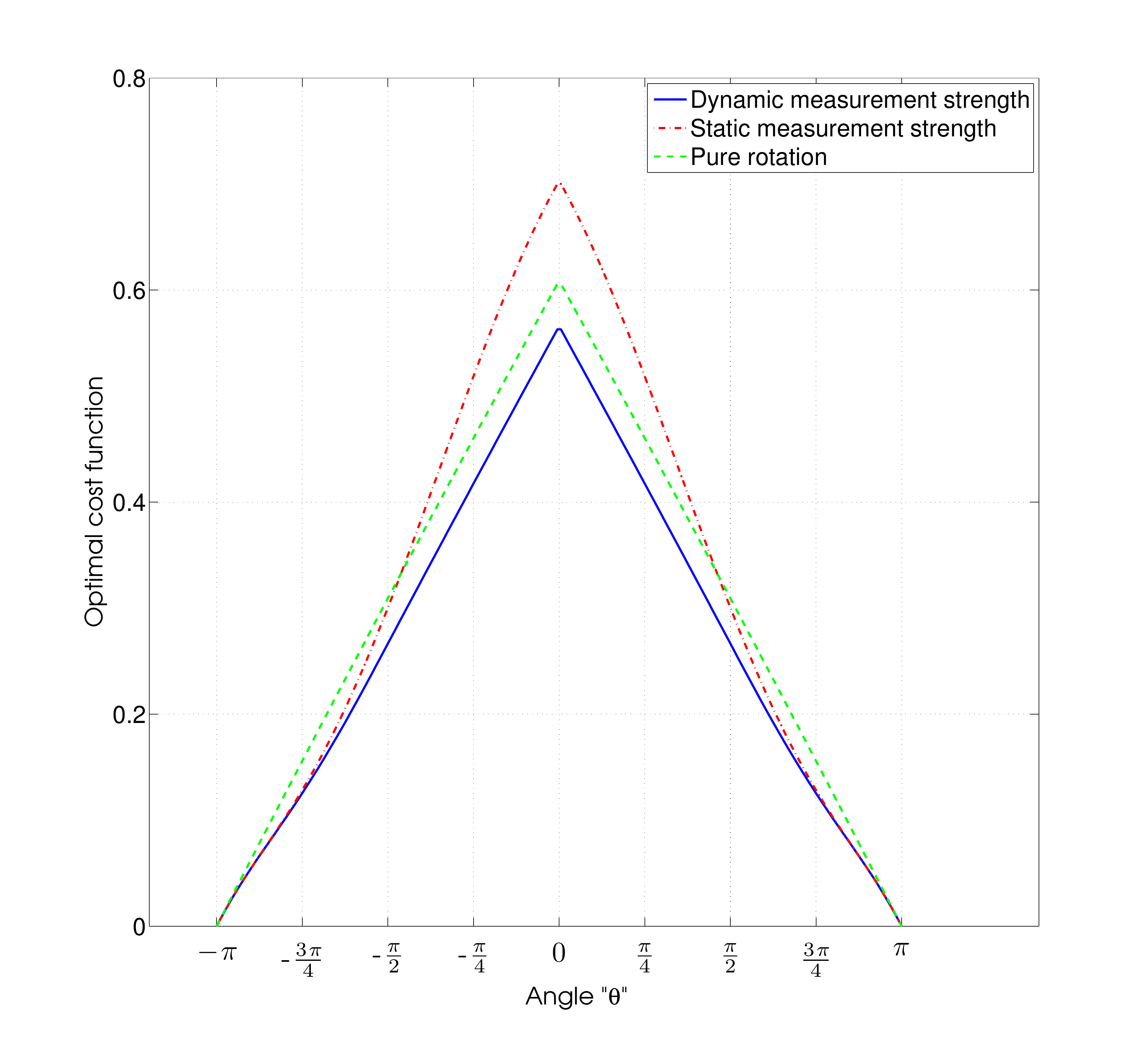}
\caption{Discounted hitting time cost function  to the target $\targetSet_{e}:=\{-\pi,  \pi \}$, starting from various possible initial states with $\controlBound = \ctrlboundsimval $, $\beta = \betasimval$ and $\Gamma = \gammasimLim$.}
\label{\figname evalueFunction}
\end{center}
\end{figure}

\begin{figure}[h!]
\label{\figname eigtoeigmeasandcontrol}
\centering
\subfigure[Optimal angular velocity control signal to   the target$\targetSet_{e}:=\{-\pi,  \pi\}$, starting from various possible initial states with $\controlBound = \ctrlboundsimval $, $\beta = \betasimval $ and $\Gamma = \gammasimLim$ .]{
\includegraphics[width=.95\hsize]{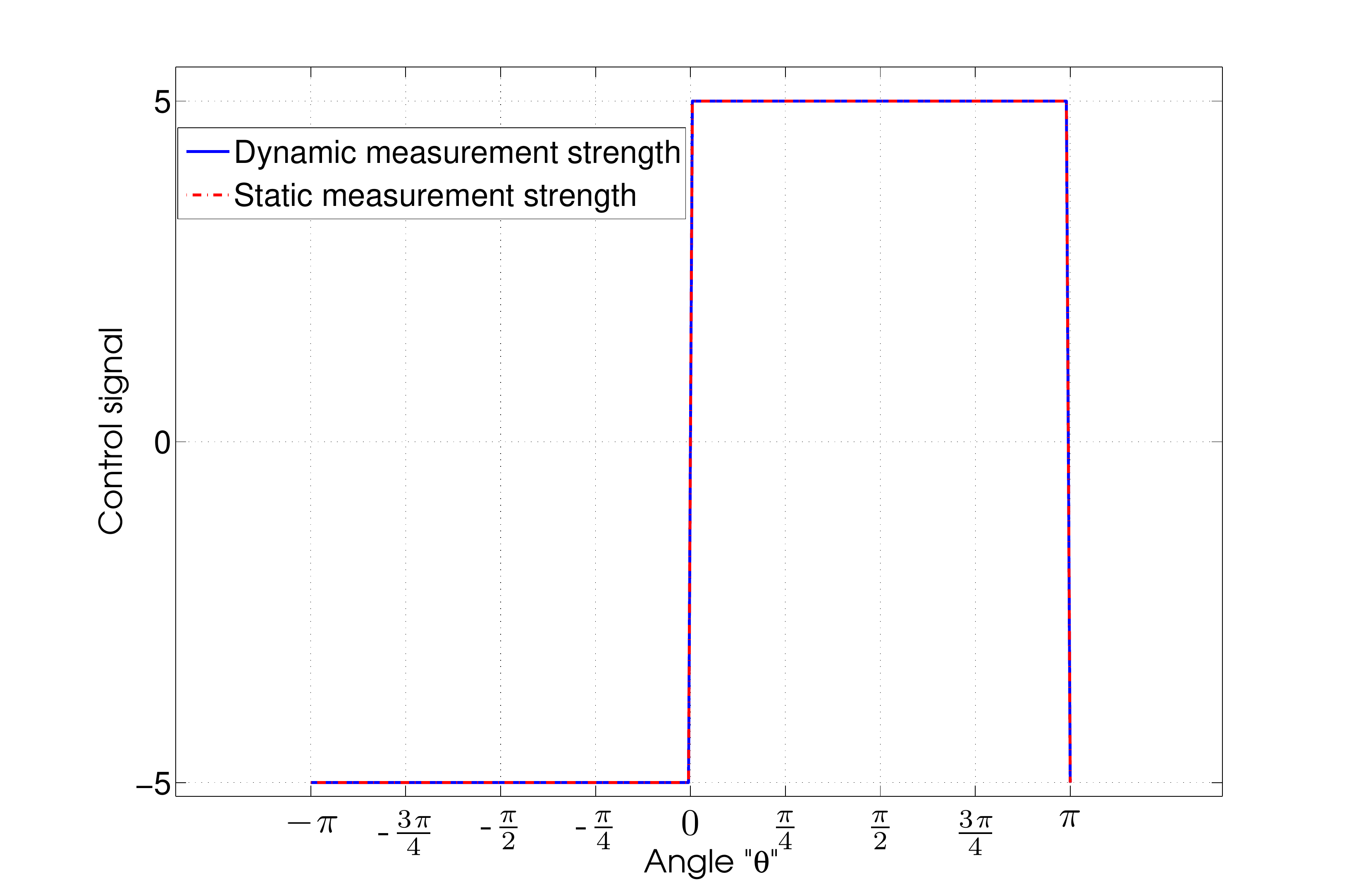}
\label{\figname controlfneigtoeigsim}
}
\subfigure[Optimal measurement strength signal to   the target $\targetSet_{e}:=\{-\pi,  \pi\}$, starting from various possible initial states with $\controlBound = \ctrlboundsimval $, $\beta = \betasimval $ and $\Gamma = \gammasimLim$.]{
\includegraphics[width=.95\hsize]{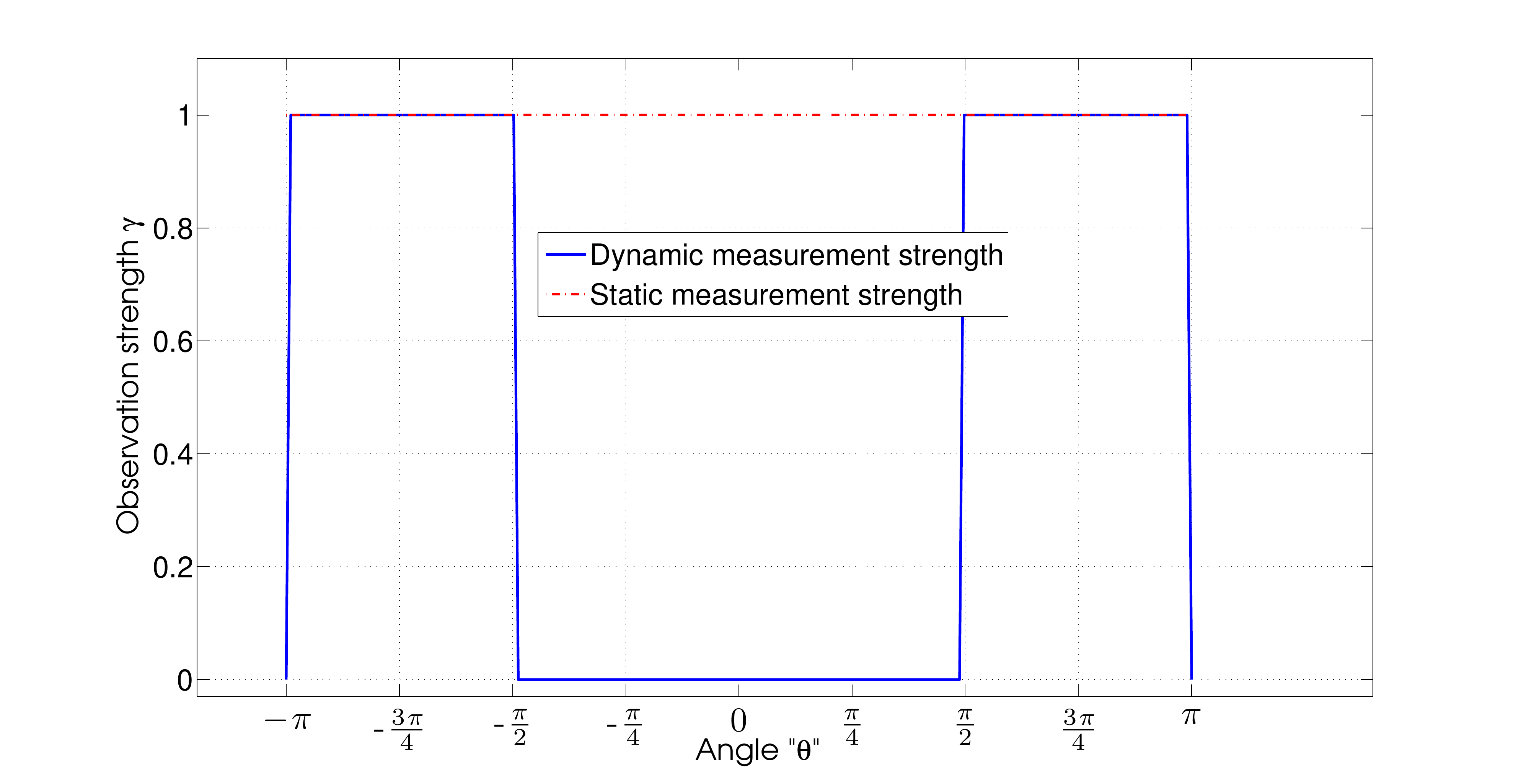}
\label{\figname obsfneigtoeig}
}
\caption{Transition between eigenstates.}
\end{figure}

In this case the states take up values from the set $G := (-\pi, +\pi)$.  Assume a control bound of $\controlBound = \ctrlboundsimval$, a discount factor of $\beta = \betasimval$ and a measurement
strength of $\Gamma= 1$. Applying the value iteration procedure, we obtain the solution to the HJB PDE \eqref{\eqnname hjbdef1} subject to
the boundary  condition of $C(\pm \pi) = 0$. The result obtained is as indicated in Fig.~\ref{\figname evalueFunction}:  the corresponding optimal control 
is as shown in Fig.~\ref{\figname controlfneigtoeigsim} and measurement strength is  indicated in Fig.~\ref{\figname obsfneigtoeig}.  Hence it turns out that the optimal control is consistent with the intuition of exercising a clockwise rotation when starting at any point to the right of the $+\ket{z}$ (up) state and  counterclockwise rotation to the left of this state.  Note that the optimal measurement strategy in this case is to turn off measurement till arriving at the state $\theta= \pi/2$.

\subsection{Optimal transition between non-eigenstates}
\begin{figure}
\begin{center}
\includegraphics[width=0.9\hsize]{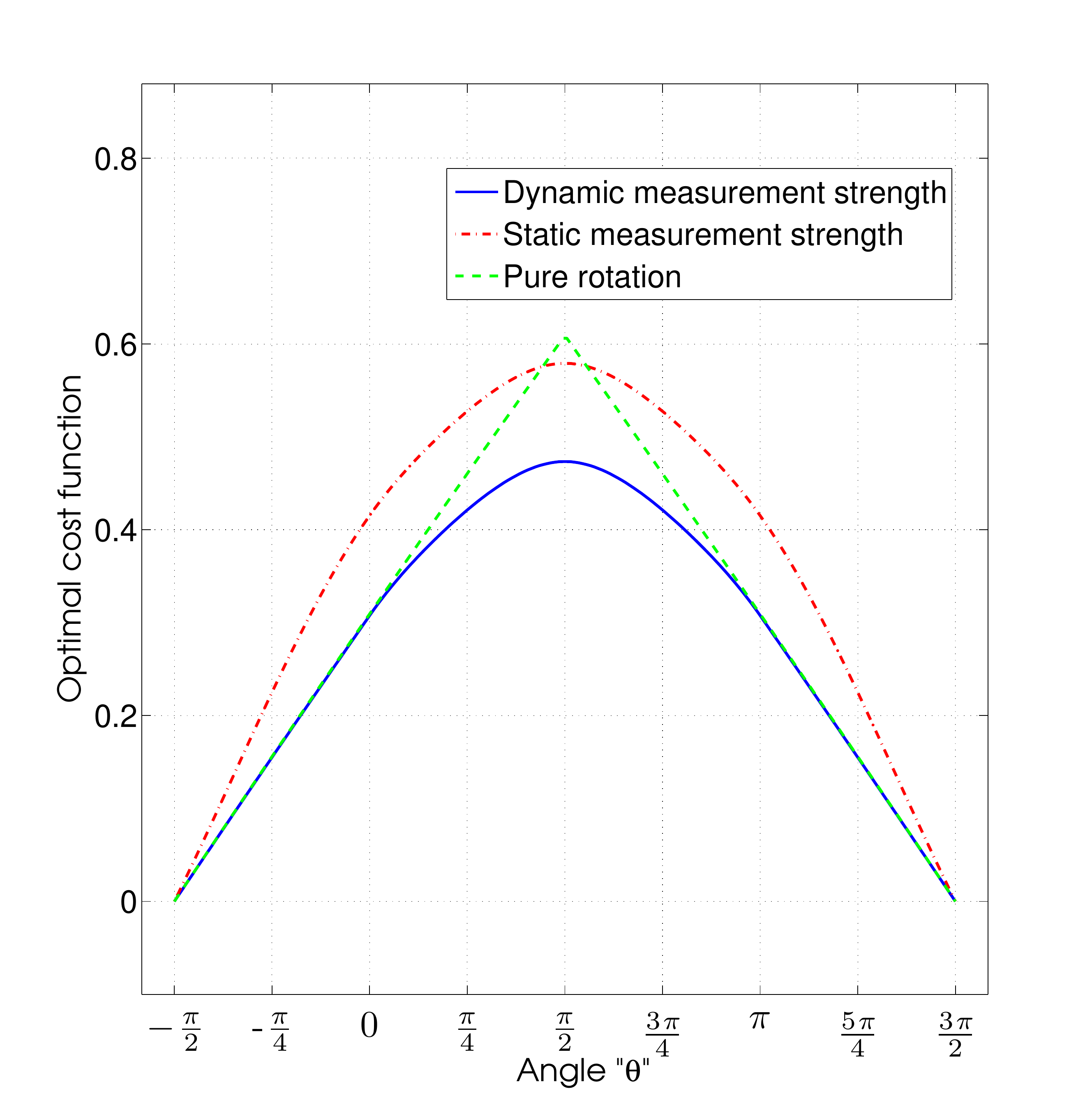}
\caption{Discounted hitting time cost function  to   the target$\targetSet_{ne}:=\{-\pi/2, 3 \pi/2\}$, starting from various possible initial states with $\controlBound = \ctrlboundsimval $, $\beta = \betasimval $ and $\Gamma = \gammasimLim$.}
\label{\figname costfnnoneigtononeigsim}
\end{center}
\end{figure}

We now study the optimal control problem of taking the initial state $\thn{0} = \pi/2$  to the target state of $\targetSet_{ne} = \{-\pi/2, 3 \pi/2\}$ as depicted in Fig.~\ref{\figname Fignoneigtononeig}. 
In this case the region of interest is $G:=(-\pi/2, 3 \pi/2)$ and the HJB  equation associated with this problem is \eqref{\eqnname hjbdef1} with a boundary condition of $C(\targetSet_{ne}) = 0$.  The value iteration approach  for this problem  yields 
the solution (ref. Fig.~\ref{\figname costfnnoneigtononeigsim}). 
This control  strategy in this case involves an angular rotation of $+\Omega$ for all states corresponding to angles  between $(\pi/2, 3 \pi/2)$ and counterclockwise control for states in  the domain $(-\pi/2, \pi/2)$.

%

\section{Interpretation}
In this section we analyze and compare the performance of the variable measurement strength approach proposed herein to two alternate control approaches - pure Hamiltonian rotation and fixed strength control.
The strategy used  as the base line for these comparisons is the case of pure rotation with no measurement. 
We now analyze the salient features of these strategies.

\subsection{Fixed measurement strength}
For the case of rotation between eigen states, from Fig.~\ref{\figname evalueFunction}  it can be seen that for angles from $[0, \pi/2)$ the fixed measurement strength strategy performs worse than the pure rotation approach; However for angles between $(\pi/2, \pi)$ the backaction from the  fixed measurement helps project the state towards the target thereby outperforming the pure rotation strategy. This is in contrast to the control between non-eigen states  $\pm \ket{x}$ where the  fixed measurement strategy performs better than the fixed rotation close to the starting point but  worse for all points further away (except at the target).  

\subsection{Dynamic measurement strength}
As can be observed from  Figures  \ref{\figname evalueFunction}, \ref{\figname costfnnoneigtononeigsim} the time for the dynamic measurement approach is always smaller than that for both:  the pure rotation strategy (in fact they are equal only at the boundary), and for the case of a static measurement. For  $ \theta \in [0,\pi/2]$  the time to reach $\theta = \pi$ is substantially different.  Hence using a dynamic control and measurement scheme shortens the hitting time to the desired state by a significant margin. In the case of the transfer between the states orthogonal to $\ket{z}$ (i.e., $\ket{x}$) we have that this strategy does better than the pure rotation strategy only for starting points from $(0, \pi)$  and equal to the rotation strategy at all other points. This is intuitive as the the dynamic measurement strategy leads to the measurement signal being  switched off for points between $(-\pi/2, 0)$ and $(\pi, 3 \pi/2)$ - leading to the use of only the maximum magnitude of the available angular rotation. 
%

\section{Conclusion and  Future Work}\label{\secname conc}
In this article we described an approach for the  time optimal rotation of a quantum two level system. The dynamic measurement and velocity control strategy led to a speedup in the hitting time compared to strategies used previously in the literature. Numerical solutions to certain example 
problems were indicated and the analysis of the solutions led to the following  interesting  avenues 
for future investigation. 

The special form of the nonlinear dynamics in the system under 
consideration, leads to a degenerate Hamilton-Jacobi-Bellman equation associated with the optimal control problem -- this necessitates  a weak (viscosity) solution interpretation for the 
solution  to these  control problems. Hence  this  remains to be investigated in greater detail. 
Another aspect of note in this problem is the fact that the optimal measurement and control strategy 
are not separable\footnote{This  is an interesting contrast to the work in \cite{gough2005hamilton} which utilizes the separation aspect of filtering and control.}; i.e., the observation strategy depends on a knowledge of the angular velocity control.  Certain aspects of this problem appear to parallel those in the well known Witsenhausen
counterexample \cite
{witsenhausen1968counterexample,castanon1978signaling,witsenhausen2010demystifying}. 
Inspired by these ideas,  we note that in the quantum control problem, one possible  viewpoint is to analyze this problem as a combination of two interrelated controllers  (i) rotational velocity control; (ii) measurement control,  
with an overall time optimal strategy depending on two controllers in a distributed control paradigm. In this arrangement, the knowledge of the state from the first controller is passed to the second controller. This potentially gives rise to a non-classical information pattern. The analysis of the impact of this on optimal control and measurement and its relationship to the foundations of decentralized control also offers a fruitful topic for future work.

%
%

{\em Aknowledgements:} The author would like to thank Joshua Combes for helpful technical discussions and comments on this manuscript.

%
%

\begin{thebibliography}{10}

\bibitem{nielsen2000qca}
M.A. Nielsen and I.L. Chuang.
\newblock {\em Quantum Computation and Quantum Information}.
\newblock Cambridge University Press, 2000.

\bibitem{higgins2007entanglement}
B.L. Higgins, DW~Berry, SD~Bartlett, H.M. Wiseman, and GJ~Pryde.
\newblock {Entanglement-free Heisenberg-limited phase estimation}.
\newblock {\em Nature}, 450(7168):393--396, 2007.

\bibitem{WisMil10}
H.M. Wiseman and G.J. Milburn.
\newblock {\em {Quantum Measurement and Control}}.
\newblock Cambridge University Press, 2009.

\bibitem{dalessandro2002ufg}
D.~D'Alessandro.
\newblock Uniform finite generation of compact lie groups.
\newblock {\em Systems and Control Letters}, 47(1):87--90, 2002.

\bibitem{schirmer2001qcu}
{\em Quantum control using Lie group decompositions}, volume~1, 2001.

\bibitem{ramakrishna2000qcd}
V.~Ramakrishna, K.L. Flores, H.~Rabitz, and R.J. Ober.
\newblock {Quantum control by decompositions of SU$(2)$}.
\newblock {\em Physical Review A}, 62(5):53409, 2000.

\bibitem{nielsen2006qcg}
M.A. Nielsen, M.R. Dowling, M.~Gu, and A.C. Doherty.
\newblock Quantum computation as geometry.
\newblock {\em Science}, 311(5764):1133--1135, 2006.

\bibitem{N.2001}
N.~Khaneja, R.~Brockett, and S.~J. Glaser.
\newblock Time optimal control in spin systems.
\newblock {\em Phys. Rev. A}, 63:032308, 2001.

\bibitem{belavkin1999measurement}
V.P Belavkin.
\newblock {Measurement, Filtering and Control in Quantum Open Dynamical
  systems}.
\newblock {\em Reports on Mathematical Physics}, 43(3):405--425, 1999.

\bibitem{wiseman2008optimality}
H.M. Wiseman and L.~Bouten.
\newblock {Optimality of feedback control strategies for qubit purification}.
\newblock {\em Quantum Information Processing}, 7(2):71--83, 2008.

\bibitem{shabani2008locally}
A.~Shabani and K.~Jacobs.
\newblock {Locally optimal control of quantum systems with strong feedback}.
\newblock {\em Physical review letters}, 101(23):230403, 2008.

\bibitem{belavkin2009dynamical}
V.P. Belavkin, A.~Negretti, and K.~M{\o}lmer.
\newblock {Dynamical programming of continuously observed quantum systems}.
\newblock {\em Physical Review A}, 79(2):22123, 2009.

\bibitem{gough2005hamilton}
J.~Gough, V.P Belavkin, and O.G Smolyanov.
\newblock {Hamilton--Jacobi--Bellman equations for quantum optimal feedback
  control}.
\newblock {\em Journal of Optics B: Quantum and Semiclassical Optics},
  7:S237--S244, 2005.

\bibitem{sjm2011optimalqbitcontrol}
Srinivas Sridharan, Masahiro Yanagisawa, and Joshua Combes.
\newblock Optimal rotation control and weak solutions for a qubit subject to
  continuous measurement.
\newblock {\em Transactions on Automatic Control, Special Issue on Quantum
  Control}, (under review).

\bibitem{brun2002}
T.~A. Brun.
\newblock A simple model of quantum trajectories.
\newblock {\em American Journal of Physics}, 70:719, 2002.

\bibitem{steck2006}
K.~Jacobs and D.~A. Steck.
\newblock {A straightforward introduction to continuous quantum measurement}.
\newblock {\em Contemporary Physics}, 47:279, 2006.

\bibitem{fleming2006cmp}
W.H. Fleming and H.M. Soner.
\newblock {\em Controlled Markov Processes and Viscosity Solutions}.
\newblock Springer Verlag, Berlin-NY, 2006.

\bibitem{sridharan2010numerical}
Srinivas Sridharan and Matthew~R. James.
\newblock {Numerical Solution of the Dynamic Programming Equation for the
  Optimal Control of Quantum Spin Systems}.
\newblock {\em Systems \& Control Letters (To appear)}, 2011.

\bibitem{bellman2003dp}
R.E. Bellman.
\newblock {\em Dynamic Programming}.
\newblock Courier Dover Publications, 2003.

\bibitem{bertsekas1995dpa}
D.P. Bertsekas.
\newblock {\em Dynamic Programming and Optimal Control}.
\newblock Athena Scientific, 1995.

\bibitem{gilbarg2001elliptic}
D.~Gilbarg and N.S. Trudinger.
\newblock {\em {Elliptic partial differential equations of second order}}.
\newblock Springer Verlag, 2001.

\bibitem{wong1985stochastic}
E.~Wong and B.~Hajek.
\newblock {\em {Stochastic processes in engineering systems. Rev. ed}}.
\newblock Springer Texts in Electrical Engineering, 1985.

\bibitem{gardiner1985handbook}
C.W. Gardiner.
\newblock {\em {Handbook of stochastic methods}}.
\newblock Springer Berlin, 1985.

\bibitem{jacobs2010stochastic}
K.~Jacobs.
\newblock {\em {Stochastic processes for physicists: understanding noisy
  systems}}.
\newblock Cambridge Univ Pr, 2010.

\bibitem{kushner1992nms}
H.J. Kushner and PG~Dupuis.
\newblock {\em Numerical Methods for Stochastic Control Problems in Continuous
  Time}.
\newblock Springer Verlag, Berlin-NY, 1992.

\bibitem{smjpra2008}
Srinivas Sridharan, Mile Gu, and Matthew~R. James.
\newblock Gate complexity using dynamic programming.
\newblock {\em Physical Review A (Atomic, Molecular, and Optical Physics)},
  78(5):052327, 2008.

\bibitem{witsenhausen1968counterexample}
H.S. Witsenhausen.
\newblock A counterexample in stochastic optimum control.
\newblock {\em SIAM Journal on Control}, 6:131, 1968.

\bibitem{castanon1978signaling}
D.A. Castanon and N.R. Sandell.
\newblock Signaling and uncertainty: a case study.
\newblock In {\em Decision and Control including the 17th Symposium on Adaptive
  Processes, 1978 IEEE Conference on}, volume~17, pages 1140--1144. IEEE, 1978.

\bibitem{witsenhausen2010demystifying}
H.S. Witsenhausen.
\newblock {Demystifying the Witsenhausen Counterexample}.
\newblock {\em IEEE Control Systems Magazine}, 1066(033X/10), 2010.

\end{thebibliography}

\end{document}